\documentclass[amsmath, amssymb, preprintnumbers, noshowpacs,aps,superscriptaddress,onecolumn]{revtex4}

\usepackage{amsmath}
\usepackage{amssymb}
\usepackage{braket}
\usepackage{graphicx}
\usepackage{verbatim}
\usepackage{hyperref}

\begin{document}
	
\title{Action Functional for a Particle with Damping}

\author{Federico de Bettin}
\affiliation{Dipartimento di Fisica e Astronomia ÒGalileo GalileiÓ,
	Universit\`a di Padova, via Marzolo 8, 35131 Padova, Italy}

\author{Alberto Cappellaro}
\email{cappellaro@pd.infn.it}
\affiliation{Dipartimento di Fisica e Astronomia ÒGalileo GalileiÓ,
	Universit\`a di Padova, via Marzolo 8, 35131 Padova, Italy}
\author{Luca Salasnich}
\affiliation{Dipartimento di Fisica e Astronomia ÒGalileo GalileiÓ,
	Universit\`a di Padova, via Marzolo 8, 35131 Padova, Italy}
\affiliation{CNR-INO, via Nello Carrara, 1 - 50019 Sesto Fiorentino, Italy}
\date{\today{}}

\begin{abstract}
	In this brief report we discuss the action functional 
	of a particle with damping, showing that it 
	can be obtained from the dissipative equation of motion through a 
	modification which makes the new dissipative equation invariant 
	for time reversal symmetry. This action functional 
	is exactly the effective action of Caldeira-Leggett model 
	but, in our approach,  
	it is derived without the assumption that the particle 
	is weakly coupled to a bath of infinite harmonic oscillators.
\end{abstract}

\maketitle
	
\section{Introduction}

The number of different systems and physical variables 
displaying damped dynamics
is vast. A dissipative equation 
of motion can be found in various models, where 
the degree of freedom undergoing a damped evolution
can be the spatial coordinate of a classical 
particle moving inside a fluid \cite{arnold}, but also, for instance,  
the phase difference of a bosonic field at a josephson 
junction \cite{eckern,zaikin,nagaosa}, or a scalar field in the theory 
of warm inflation \cite{morikawa,paz,calzetta}. 
Obviously, the issue of dynamical evolution displaying dissipation has been 
analyzed also in the quantum realm, by making use of a great variety of techniques,
spanning from the quasiclassical Langevin equation and stochastic modelling
\cite{schmid,grabert} to a refined Bogoliubov-like approach for the motion
of an impurity through a Bose-Einstein condensated \cite{lampo}.
 
The presence of a dissipative term in the equation of motion 
makes the formulation of a variational principle and the derivation of 
an action functional quite problematic. On the other hand, the knowledge of the 
effective action of a particle with damping is crucial 
to study the role of dissipation on the quantum tunneling of 
the particle between two local minima of its confining 
potential \cite{ingold}. 

There are various approaches to the construction of an action fuctional 
for a particle with damping. In 1931 Bateman \cite{bateman1931}
derived it by exploiting the variational principle with a Lagrangian involving 
the coordinates of the particle of interest and an additional 
degree of freedom. The price to pay in this doubling of the variables 
is a complicated expression for the kinetic energy, 
which does not have a simple quadratic form. In 1941 another approach 
was suggested by Caldirola \cite{caldirola1941}, who wrote an explicitly 
time-dependent Lagrangian, whose Euler-Lagrange equation gives exactly 
the equation of motion with a dissipative term. In 1981 Caldeira and 
Leggett \cite{leggett1981} considered a particle weakly coupled 
to the environment, modelled by a large number of harmonic oscillators. 
Integrating out the degrees of freedom of the harmonic oscillators,   
they found an effective action for the particle with dissipation 
and then they used it to calculate the effect of the damping coefficient 
on the quantum tunneling rate of the particle 
\cite{leggett1981,leggett1982,leggett1984}. 
This framework started a deep theoretical effort devoted to understand 
quantum dynamics in a dissipative environment with its 
related features such as the localization transition \cite{schmid1983} and 
the diffusion in periodic potential \cite{weiss,zwerger,friedman}.

In this review paper we discuss a shortcut of the treatment made by Caldeira 
and Leggett \cite{leggett1981,leggett1984}. We show that the Caldeira-Leggett 
effective action can be obtained, without the assumption of an 
environmental bath, directly from the dissipative equation of 
motion through a modification 
which makes the new dissipative equation of motion invariant 
for time reversal symmetry. Our approach is somehow similar to the 
one recently proposed by Floerchinger \cite{floerchinger2016} 
via analytic continuation. The paper is organized as follows. 
First, in Section 2 we briefly review the quantum 
tunneling probability of the particle on the basis of the saddle-point 
approximation of the path integral with imaginary time. Then, in Section 3 
we discuss the Caldeira-Leggett approach, where the effective 
action of a particle with damping 
is obtained assuming that the particle is weakly coupled to a bath 
of harmonic oscillators. Finally, in Section 4 we show this effective action 
can be obtained, without the assumption of an environmental bath, 
directly from a modified dissipative equation of motion. 

\section{Path Integral Formulation of Quantum Tunneling} 

We consider a single particle of mass $m$ and coordinate $q(t)$ 
subject to an external potential $V(q)$ 
with a metastable local minimum at $q_I$. 

In this Section we discuss the quantum tunneling of the particle 
from the metastable local minimum in the absence of dissipation. 
The transition amplitude for the particle located at the metastable minimum 
$q_I$ at time $t_F$ to propagate to the position $q_F$ at time $t_F$ 
can be computed using the Feynman path integral
\begin{equation}
\braket{q_F,t_F|q_I,t_I}=\int_{q(t_I)=q_I}^{q(t_F)=q_F} D[q(t)] \, 
e^{\frac{i}{\hbar}S[q(t)]} \; , \label{eq:feynmanprop}
\end{equation}
where 
\begin{equation}
S[q(t)]=\int_{t_I}^{t_F} dt \, L(q(t)) 
\end{equation}
is the action functional of the system and $L(q(t))$ is its Lagrangian, 
which has the form
\begin{equation}
L(q)=\frac{1}{2}m\dot{q}^2-V(q) \; . \label{eq:lagrange}
\end{equation}

To obtain an analytical approximation of \eqref{eq:feynmanprop} one would like 
to use the saddle-point approximation, expanding the action around 
the ``classical trajectory'', that is the one minimizing the action. 
The problem is that, without the help of some external energy, 
there is no classical trajectory starting from $q(t_I)=q_I$ 
with $\dot{q}(t_I)=0$ and ending at $q(t_F)=q_F$. 
What we can do is to evaluate the imaginary time path integral, obtained by 
performing the change of variable $t=-i\tau$. Such operation is simply a 
$\pi/2$ rotation in the complex $t$ plane which, given that no poles 
are encountered during the rotation, does not change the results of the 
integral even if the time integration is performed in the real domain after 
the change of variables. Thus, Eq. \eqref{eq:feynmanprop} becomes
\begin{equation}
\braket{q_F,\tau_F|q_I,\tau_I}=\int_{q(\tau_I)=q_I}^{q(\tau_F)=q_F} 
D[q(\tau)] \, e^{-\frac{1}{\hbar}S_E[q(\tau)]} \; , 
\end{equation}
where $S_E[q(\tau)]$ is the so-called Euclidean action 
\begin{equation}
S_E[q(\tau)]=\int_{-\infty}^{+\infty}d\tau \, L_E(q(\tau)) \; ,
\end{equation}
and $L_E(q(\tau))$ the corresponding Euclidean Lagrangian which, in the usual 
conservative case \eqref{eq:lagrange} is simply the Lagrangian with inverted 
potential $V(q)\rightarrow-V(q)$
\begin{equation}
L_E(q)=\frac{1}{2}m\dot{q}^2+V(q) \; . 
\end{equation}
Because of this, the minima of $V(q)$ behave like maxima in the Euclidean 
action and so there exists a trajectory which minimizes $S_E[q(\tau)]$ 
starting from $q_I$ and ending at $q_F$.

Then, to calculate an approximate expression for the transition amplitude, 
one finds the ``imaginary-time classical trajectory'' $\bar{q}(\tau)$, which minimizes 
the Euclidean action  with the chosen boundary 
conditions, and one eventually gets 
\begin{equation}
\braket{q_F,\tau_F|q_I,\tau_I} \simeq A \, e^{-S_E[\bar{q}(\tau)]/\hbar} \; , 
\label{eq:decayrate}
\end{equation}
where 
\begin{equation}
A=\Big(\frac{S_E[\bar{q}(\tau)]}{2\pi\hbar}\Big)^{\frac{1}{2}} \, \Big(\frac{\det
\big(-\partial^2_\tau+V''(q_I)\big)}
{\det'\big(-\partial^2_\tau+V''(\bar{q})\big)}
\Big)^{\frac{1}{2}} \; 
\end{equation}
with $\det'(\cdot)$ the determinant computed by 
excluding the null eigenvalues \cite{coleman1977,wen}.

Our interest is to study how a particle subject to friction can escape from 
a metastable state driven by quantum tunneling. The action with Lagrangian 
\eqref{eq:lagrange}, in such case, will not be useful, as it cannot describe 
a dissipative system. We want to study how the presence of friction influences 
the form of \eqref{eq:decayrate}, and to do so we need to know the form of 
the Euclidean action of a dissipative system. It can be derived by means of
the Caldeira-Leggett model, which is presented in the following Section.
	
\section{The Caldeira-Leggett Model}

The Caldeira-Leggett model describes the motion of a particle in one dimension 
in a heat bath made of $N$ decoupled harmonic oscillators, each with 
a characteristic frequency $\omega_j$, with 
$j\in\{1,2,\dots, N\}$; their coordinates will be labeled as $x_j(t)$. 
The particle, described by the coordinate $q(t)$, is 
subject to an external potential $V(q)$ and coupled to the 
$j$-$th$ harmonic oscillator via the coupling constant $g_j$.  
	
The Lagrangian describing this system is 
\begin{equation}
L=\frac{1}{2}m\dot{q}^2-V(q)+\sum_{j=1}^N\left(\frac{1}{2}m_j\dot{x}_j^2-
\frac{1}{2}m_j\omega_j^2x_j^2
\right)+q\sum_{j=1}^Ng_jx_j-q^2\sum_{j=1}^N\frac{g_j^2}{2m_j\omega_j^2} \; . 
\label{eq:Caldeira_leggett_lagrangian}
\end{equation}
The last term is simply a counterterm not depending on the oscillator coordinates. 
The physical reason of the introduction of such term is to let the minimum 
of the Hamiltonian, and thus of the energy, corresponds to the minimum 
of the external potential $V(q)$.
	
Given Eq. \eqref{eq:Caldeira_leggett_lagrangian}, the action immediately reads
\begin{equation}
S[q(t)]=\int_{t_I}^{t_F} dt \, \left(\frac{1}{2}m\dot{q^2}
-\frac{\partial V(q)}{\partial q}
+\sum_{j=1}^N\frac{1}{2}m_j\dot{x}_j^2-\sum_{j=1}^N\frac{m_j}{2}
\left(\omega_jx_j-\frac{g_j}{m_j\omega_j}
q\right)^2\right) \; , \label{eq:action}
\end{equation}
while the transition amplitude for the particle to propagate from 
position $q_I$ at time $t_I$ to 
position $q_F$ at time $t_F$ and for the $j-th$ harmonic 
oscillator to propagate from coordinate $x_{j,I}$ at time $t_I$ 
to $x_{j,F}$ at $t_F$ can be written as
\begin{equation}
\braket{q_F,\{x_{j,F}\},t_F|q_I,\{x_{j,I}\},t_I}=
\int_{q(t_I)=q_I}^{q(t_F)=q_F} D[q] \, 
\Big(\prod_{j=1}^N\int_{x_j(t_I)=x_{j,I}}^{x_{j}(t_F)=x_{j,F}} 
D[x_j]\Big) \, e^{\frac{i}{\hbar}S[q;\{x_j\}]} \; . 
\label{eq:pathintegral}
\end{equation}
	
The degrees of freedom of the environment are of no actual interest, 
and from \eqref{eq:pathintegral} we would like to obtain a theory involving 
 only the  degrees of freedom of the particle, as suggested by 
Feynman and Vernon \cite{feynman}.
We will build an effective theory for the system, with effective action 
$S^{eff}[q(t)]$, by integrating out all 
the degrees of freedom of the environment, such that
\begin{equation}
\Big(\prod_{j=1}^N\int 
D[x_j]\Big) \, e^{\frac{i}{\hbar}S[q;\{x_j\}]} = e^{\frac{i}{\hbar}S^{eff}[q(t)]} \; , 
\end{equation}
where the initial conditions $\{x_{j,I}\}$ and the 
final conditions $\{x_{j,F}\}$ can assume any real value. 
Luckily, this is a doable task thanks to the fact that in the action the 
highest polinomial degree in $x_j$ is $x_j^2$ and that the coupling with 
the particle is bilinear.

The path integral over the coordinates of the environment can be decoupled 
from the one over the paths of the particle of interest. 
This procedure is discussed in detail in Refs. \cite{ingold,wen}. 
We then obtain an effective action 
\begin{equation}
\begin{split}
S^{eff}[q(t)]=\int_{t_I}^{t_F} dt \, 
\Big(\frac{m}{2}\dot{q}^2-V(q)-q^2\sum_{j=1}^N\frac{g_j^2}{2m_j\omega_j^2}
\Big)+ \\
+i\sum_{j=1}^N\frac{g_j^2}{4m_j\omega_j}\int_{t_I}^{t_F} dt 
\int_{-\infty}^{+\infty}dt' \, q(t) \, q(t') \, e^{-i\omega_j|t-t'|} \; ,
\end{split}
\end{equation}
which can also be re-written as 
\begin{equation}
S^{eff}[q(t)]=\int_{t_I}^{t_F} dt\left(\frac{m}{2}\dot{q}^2-V(q)+ 
\frac{1}{4}\int_{-\infty}^{+\infty} dt' \, K(t-t')\, 
\Big(q(t)-q(t')\Big)^2\right) \; , 
\label{eq:effectiveaction}
\end{equation}
where
\begin{equation}
K(t-t')=-i\sum_{j=1}^N\frac{g_j^2}{2m_j\omega_j}e^{-i\omega_j|t-t'|}
\xrightarrow[N\rightarrow+\infty]{\text{ }} -i\int_0^{+\infty} 
d\omega\frac{g^2(\omega)}{2m\omega}n(\omega)e^{-i\omega|t-t'|} \; , 
\label{eq:Greenfunction}
\end{equation}
assuming that the spectrum of frequencies of the bath is continuous 
with $n(\omega)$ the number of harmonic oscillators with frequency 
$\omega$ and that the masses of the oscillators are all the same.
We can now use Eq. \eqref{eq:effectiveaction} to
express the particle propagator
 as
\begin{equation}
\braket{q_F,t_F|q_I,t_I}=\int_{q(t_I)=q_I}^{q(t_F)=q_F} 
D[q] \, e^{\frac{i}{\hbar}S^{eff}[q(t)]} \; . 
\label{eq:efftheory}
\end{equation}
The equation of motion obtained by extremizing 
\eqref{eq:effectiveaction} is (see also \cite{wen}) 
\begin{equation}
-m\ddot{q}(t)-\frac{\partial V(q)}{\partial q}+\int_{-\infty}^{+\infty} 
dt'K(|t-t'|)q(t)-\int_{-\infty}^{+\infty} dt'K(|t'-t|)q(t')= 0 \; ,
\end{equation}
which in frequency space reads
\begin{equation}
- m\omega^2{\tilde q}(\omega) + ({\tilde K}(\omega)-{\tilde K}(0)){\tilde q}
(\omega)+{\cal F} \Big[\frac{\partial V}{\partial q}\Big](\omega) = 0 \; , 
\label{eq:equationofmot}
\end{equation}
where by ${\cal F}[\cdot]$ we denote the Fourier transform. 
For the sake of simplicity, we now assume 
the following low frequency behaviour of the kernel:
\begin{equation}
{\tilde K}(\omega)=-i\gamma|\omega|. \label{eq:definitiongammam}
\end{equation}
Equation \eqref{eq:equationofmot}, then, becomes
\begin{equation}
-m\omega^2{\tilde q}(\omega)-i|\omega|\gamma {\tilde q}(\omega) = - 
{\cal F}\Big[\frac{\partial V}{\partial 
	q}\Big](\omega) \; . 
\label{eq:fouriereqofmot}
\end{equation}
By taking \eqref{eq:fouriereqofmot} to real-time space the equation becomes 
\begin{equation}
m {\ddot q} + {i\over \pi} \gamma 
\int_{-\infty}^{+\infty} dt' {q(t') \over (t-t')^2} = 
- \frac{\partial V(q)}{\partial q}  \; , 
\label{eq:effectiveeqofmot}
\end{equation}
which describes a particle of mass $m$ subject to a nonlocal 
friction, with damping coefficient $\gamma$. A detailed 
discussion of this real-time equation and the associated real-time 
action fuctional (\ref{eq:effectiveaction}) can be found in Ref. \cite{wen}. 

\section{Direct Derivation of the Action Functional for 
a Particle with Damping}
	
The discussion of the last part of the previous section 
is useful for our scope of deriving the action functional for a particle 
subject to friction directly from its equation of motion, 
without the need for the introduction of a bath made of harmonic oscillators.
	
Let us consider a particle of mass $m$ and 
coordinate $q(t)$ in the presence of an external conservative force
\begin{equation}
F_c=-\frac{\partial V(q)}{\partial q}
\end{equation}
with $V(q)$ the corresponding potential energy, and also under the effect 
of a dissipative force
\begin{equation}
F_d=-\gamma\dot{q} \; ,
\end{equation}
with $\gamma>0$ the damping coefficient. The equation of motion for the 
particle is given by
\begin{equation}
m\ddot{q}+\gamma\dot{q}=-\frac{\partial V(q)}{\partial q} \; . 
\label{eq:S1}
\end{equation}
The Newton equation is clearly not invariant for time reversal 
$(t\rightarrow-t)$ due to the dissipative term, which contains a first 
order time derivative.
	
The Fourier transform of Eq. \eqref{eq:S1} reads
\begin{equation}
-m\omega^2{\tilde q}(\omega)-i\gamma \omega {\tilde q}(\omega)=
-{\cal F}\Big[\frac{\partial V}{\partial q}\Big](\omega) \; , 
\label{eq:effectiveeqofmotfourier}
\end{equation}
where ${\tilde q}(\omega)={\cal F}[q(t)](\omega)$ 
is the Fourier transform of $q(t)$ and $i$ is the imaginary unit. 
This equation is clearly not invariant under frequency 
reversal $(\omega\rightarrow-\omega)$ due to the dissipative term, 
which has a linear dependence with respect to the frequency $\omega$.
	
Thus, we have seen that the absence of time-reversal 
symmetry implies the absence of frequency-reversal symmetry, and viceversa. 
The frequency reversal symmetry can be restored modifying 
Eq. \eqref{eq:effectiveeqofmotfourier} as follows
\begin{equation}
-m\omega^2{\tilde q}(\omega)-i\gamma|\omega|{\tilde q}(\omega)=
-{\cal F}\Big[\frac{\partial V}{\partial q}\Big](\omega) \; ,
\end{equation}
where in the dissipative term we substituted $\omega$ with $|\omega|$. 
This equation is clearly equal to Eq. (\ref{eq:fouriereqofmot}) 
of the previous Section. 

The inverse Fourier transform of this modified equation gives
\begin{equation}
m\ddot{q}(t)+\int_{-\infty}^{+\infty} dt'K(t-t')q(t')=
-\frac{\partial V(q)}{\partial q} \; , \label{eq:S2}
\end{equation}
where
\begin{equation}
K(t-t')=i\frac{\gamma}{\pi(t-t')^2}
\end{equation}
is the nonlocal kernel of our modified Newton equation \eqref{eq:S2} in 
the time domain. Unfortunately this equation, which is 
exactly Eq. (\ref{eq:effectiveeqofmot}) of the previous Section, 
depends explicitly on the imaginary units $i$ and this means 
that the coordinate $q(t)$ must be a complex number evolving in real time. 
Quite formally, Eq. \eqref{eq:S2} can be seen as the 
Euler-Lagrange equation of this complex and nonlocal action functional
\begin{equation}
S=\int_{t_I}^{t_F} dt \, 
\Big(\frac{m}{2}\dot{q}^2-V(q)\Big)+\frac{1}{4}\int_{t_I}^{t_F} 
dt\int_{-\infty}^{+\infty} dt' \, K(t-t') \, \big(q(t)-q(t')\big)^2 \; ,  
\label{eq:S3}
\end{equation}
that is indeed the Caldeira-Leggett effective action 
(\ref{eq:effectiveaction}) we have obtained in the previous Section,  
integrating out the degrees of freedom of the environmental bath. 

Performing a Wick rotation of time, i.e. 
setting $t=-i\tau$, this action functional can be written as 
\begin{equation}
S=i\, S_E,
\end{equation}
where
\begin{equation}
S_E=\int_{\tau_I}^{\tau_F} d\tau \, \left( \frac{m}{2}\dot{q}^2 +V(q) \right) 
+\frac{1}{4} \int_{\tau_I}^{\tau_F} d\tau \int_{-\infty}^{+\infty}
d\tau' \, K_E(\tau-\tau')\big(q(\tau)-q(\tau')\big)^2 \;  \label{eq:S4}
\end{equation}
is the Euclidean action, namely the action with imaginary time $\tau$, and 
\begin{equation}
K_E(\tau-\tau')=-\frac{1}{\pi}\frac{\gamma}{(\tau-\tau')^2} \label{eq:S5}
\end{equation}
is the Euclidean nonlocal kernel. It is important to stress that, 
contrary to the action $S$, the 
Euclidean action $S_E$ can be considered a real functional, 
assuming that the coordinate $q(\tau)$ 
is a real number evolving in imaginary time $\tau$.
	
We can use the action functionals \eqref{eq:S3} and $\eqref{eq:S4}$ 
to determine the probability amplitude that the particle of our system 
located at $q_I$ at $\tau_I$ arrives in the position 
$q_F$ at time $\tau_F$. This is given by
\begin{equation}
\braket{q_F,\tau_F|q_I,\tau_I}=\int_{q(\tau_I)=q_I}^{q(\tau_F)=q_F} 
D[q(\tau)]\, e^{-\frac{1}{\hbar}S_E[q(\tau)]} \; . 
\label{eq:S6}
\end{equation}
This formula, with the Euclidean action $S_E[q(\tau)]$ given by 
Eq. \eqref{eq:S4} and the dissipative kernel $K_E(\tau-\tau')$ given by 
Eq. \eqref{eq:S5}, 
is exactly the one used by Caldeira and Leggett \cite{leggett1981} 
to find the effect of dissipation on the tunneling 
probability between two local minima $q_I$ and $q_F$ of the potential $V(q)$.
	
There is, however, a remarkable difference between our 
approach and the one of Caldeira and Leggett in 
\cite{leggett1984}: we have derived Eqs. \eqref{eq:S4}, \eqref{eq:S5} 
and \eqref{eq:S6} directly from the dissipative 
equation of motion \eqref{eq:S1}, 
while Caldeira and Leggett derived these equations starting from the 
Euclidean Lagrangian of the particle  
coupled to a bath of harmonic oscillators.
	
The effective action we have derived can then be used to compute 
directly \eqref{eq:S6}, using the approximation \eqref{eq:decayrate}. 
With such action, though, $A$ and $S_E[\bar{q}(\tau)]$ will have to depend on 
$\gamma$, too. In particular, the coefficient $A$, 
following \cite{leggett1984}, becomes
\begin{equation}
A=\Big(\frac{S_E[\bar{q}(\tau)]}{\pi\hbar}\Big)^{\frac{1}{2}}\, 
\Big(\frac{\det\big(-\partial^2_\tau+K_E(\tau-\tau')+V''(q_I)\big)}
{\det'\big(-\partial^2_\tau+K_E(\tau-\tau')+ 
V''(\bar{q})\big)}\Big)^\frac{1}{2} \; ,
\end{equation}
 As reported in \cite{leggett1984}, 
the contribution of the friction coefficient to $S_E[\bar{q}(\tau)]$ is positive, so that 
friction always tends to suppress the tunneling rate.

At this point, we have to consider a broader point of view on dissipative processes and 
the dynamical evolution of a quantum system coupled to the external environment. 
Certainly, a vast literature \cite{wen,kamenev-book,berges-review} 
has clarified that the usual Feynman path integration is not suited to deal, in principle,
with dissipation and, in general, with non-equilibrium dynamics. In order to develop
a meaningful microscopic approach, the Schwinger-Keldysh closed time path integral appears to be 
a more reliable framework. Unfortunately, it is immediate to realize that the price to pay is
a much more complex formalism than the one outlined in this paper. 

However, for a wide range of problems it is possible to recover a Feynman formulation in terms 
of an effective action such as Eq. \eqref{eq:S4}. For instance, this is the case for
the quantum tunneling in a dissipative environment or the transition to a localized state
for a particle moving in a quasiperiodic potential \cite{friedman}. In these situations,
we are not interested to the full quantum dynamical evolution, but we actually restrict ourselves
to study fluctuations around an equilibrium state \cite{floerchinger2016}.  
Indeed, even when there is thermal equilibrium between the system and its environment, 
fluctuations are still present and may crucially affect correlation functions such as the position
one \cite{ingold}. 
In order to compute these important quantities rather than the dissipative equation of motion, 
it is fundamental to have a well-defined effective action
with time-reversal invariance, such as the one in Eq. \eqref{eq:S4}.

\section{Conclusions}

In conclusion, we have reviewed different approaches to the derivation 
of an action functional for a particle with damping and its crucial 
role on quantum tunneling. In Section 4 we have also proposed a slightly 
new approach by changing the dissipative equation of motion 
of a particle to make it invariant for time reversal symmetry. 
This modified equation of motion is nonlocal and complex, 
and it can be considered as the 
Euler-Lagrange equation of a nonlocal action functional. 
We have shown that this action functional is exactly 
the one derived by Caldeira and Leggett to study the effect 
of dissipation on the quantum tunneling of the particle. 
We stress again that, contrary to the Caldeira-Leggett approach, 
our action functional has been derived without 
the assumption that the particle is weakly coupled 
to a bath of infinite harmonic oscillators. 

In the end, it is worth remembering, as stated in the introduction,
that the theoretical framework outlined in this review can be effectively
used to deal with a broader class of problems, besides the modelling of dissipative quantum
tunnelling. For instance, within cold atoms experiments, it is possible to engineer
periodic or disordered potentials with an exquisite control over their characteristic parameters
and the coupling with the external environment \cite{bloch2008}. As a consequence, this
has sprung a renewed effort to understand the of a quantum system towards a localized state
\cite{friedman,huse2015,billy2008,luschen2017}. While a full understanding of the non-equilibrium
quantum dynamics may require more refined functional approaches \cite{kamenev-book}, 
it has been shown that, by using the Feynman formulation of the path integral,
one can understand this transition in great detail \cite{friedman}, where all the relevant
physical information are basically encoded in the kernel $K_E(\tau)$ defined in Eq. 
\eqref{eq:S5}.

\end{document}